\documentclass[aps,prb,showpacs,citeautoscript,twocolumn]{revtex4}

\usepackage{graphicx}
\usepackage{amsmath}
\usepackage{amssymb}

\newcommand{\bea}{\begin{eqnarray*}}
\newcommand{\eea}{\end{eqnarray*}}
\newcommand{\bne}{\begin{equation*}}
\newcommand{\ede}{\end{equation*}}

\newcommand{\bnen}{\begin{equation}}
\newcommand{\eden}{\end{equation}}
\newcommand{\bean}{\begin{eqnarray}}
\newcommand{\eean}{\end{eqnarray}}
\newcommand{\bnsn}{\begin{subequations}}
\newcommand{\edsn}{\end{subequations}}

\newcommand{\bna}{\begin{array}}
\newcommand{\eda}{\end{array}}
\newcommand{\bnm}{\begin{enumerate}}
\newcommand{\edm}{\end{enumerate}}

\renewcommand{\vec}[1]{\text{\boldmath{$ #1 $}}}
\begin{document}

\title{Hyperfine-induced valley mixing and the spin-valley blockade in carbon-based quantum dots}

\author{Andr\'as P\'alyi}

\affiliation{Department of Physics, University of Konstanz, D-78457 Konstanz, Germany}

\author{Guido Burkard}
\affiliation{Department of Physics, University of Konstanz, D-78457 Konstanz, Germany}

\date{\today}

\begin{abstract}
Hyperfine interaction (HFI) in carbon nanotube and graphene quantum dots is due to the presence of $^{13}$C atoms.
We theoretically show that in these structures the short-range nature of the HFI 
gives rise to a coupling between the valley degree of freedom of the electron and
the nuclear spin, in addition to the usual electron spin-nuclear spin coupling.
We predict that this property of the HFI affects the
Pauli blockade transport in carbon-based double quantum dots.
In particular, we show that transport is blocked only if both the spin and the valley degeneracies
of the quantum dot levels are lifted, e.g., by an appropriately oriented magnetic field.
The blockade is caused by four ``supertriplet'' states
in the (1,1) charge configuration.

%We predict that this property of the hyperfine interaction results in peculiar features in the
%Pauli blockade transport in carbon-based double quantum dots.

\end{abstract}
%73.63.Kv Quantum dots 
%73.63.Fg Nanotubes 
%73.23.Hk Coulomb blockade; single-electron tunneling 
%31.30.Gs Hyperfine interactions and isotope effects

\maketitle

%===

%\emph{Introduction.}
In the past decade, fundamental steps have been made towards the realization of 
quantum information
processing, including isolation, manipulation, and readout of single electron spins in 
the solid state \cite{Hanson-rmp}.
However, the majority of the existing quantum dot (QD) spin qubits is fabricated
in material systems, where hyperfine interaction (HFI) with nuclear spins 
limits the device performance via spin decoherence.
Carbon structures, such as carbon nanotubes (CNTs) or graphene,
are expected to have 
weak HFI, due to the small 1\% natural abundance
of spin-carrying $^{13}$C nuclei. 
This expectation has motivated intensive theoretical 
investigation \cite{Trauzettel-spinqubitsingraphene,Recher-grapheneqds,Bulaev-socincntdots,Fischer-cnt} and
the experimental realization of QDs in carbon nanostructures \cite{
Graber-cntdoubledots,Kuemmeth-spinorbit-in-cnts,Buitelaar-spinblockade,Churchill-13cntprl,Churchill-cntspinblockade,
Steele-cntdqd,Ponomarenko-diracbilliard,Stampfer-grapheneqd,Molitor-graphenedqd,Liu-coulombblockade-graphene}.
A further perspective of carbon-based quantum information processing 
has been opened by proposals suggesting to utilize the valley degree of freedom of
the delocalized electrons as a qubit \cite{Rycerz-valleytronics,Recher-graphenering}.
Relaxation and decoherence mechanisms of these valley-qubits are yet to be explored.
One possible source of those is short-range disorder, which is known 
to couple the two different valley states \cite{ando-aob1}. 

Double quantum dot (DQD) structures in the two-electron regime 
are particularly well suited for studying the effects of HFI
\cite{Ono-spinblockade,Koppens-spinblockade,Jouravlev-spinblockade,Petta-science,Coish-doubledot}.
In the so-called Pauli blockade regime, the measurement of the 
direct current through a serially coupled GaAs DQD, as a function of the 
external magnetic field, has been used to infer the hyperfine energy scale
\cite{Koppens-spinblockade,Jouravlev-spinblockade}.
Another experiment in GaAs DQDs showed that HFI 
can be utilized to perform coherent rotations between the states
of a singlet-triplet qubit and, at the same time, acts as a source
of decoherence \cite{Petta-science}.

The effect of the HFI in carbon
is most pronounced in fully $^{13}$C-enriched samples
\cite{Churchill-cntspinblockade,Churchill-13cntprl,
Yazyev-graphenehf,Braunecker-nuclearmagnetism,Dora-graphenenmr,Fischer-cnt}.
Such nanotube DQD devices have been used recently to estimate the 
%characteristic 
energy scale of the atomic HFI as $\sim 100\mu$eV,
using transport \cite{Churchill-cntspinblockade} and 
singlet-triplet dephasing time \cite{Churchill-13cntprl} measurements.
In contrast, theory predicts an atomic hyperfine energy scale
$\sim 1\mu$eV \cite{Yazyev-graphenehf,Fischer-cnt}. This discrepancy between theory and experiment, 
together with unexplained features of the dephasing time
measurements of Ref. \onlinecite{Churchill-13cntprl}, 
show that 
additional
theoretical efforts
have to be made to gain a complete understanding of the role 
%played by 
of 
HFI in carbon-based QDs.

Here, we study the influence of the $^{13}$C nuclear spins 
on the spin and valley degrees of freedom of the electrons
in carbon-based QDs. 
In particular, we derive the $4\times 4$ Hamiltonian describing the
effect of HFI on a single fourfold (spin and valley) degenerate QD energy level.
We find that due to the short-range nature of the HFI,
it couples the nuclear spins not only with the spin, but also with the 
valley degree of freedom of the electron. 
The effective hyperfine Hamiltonian can be expressed as
\bnen
\label{eq:hfsimple}
H_{\rm hf} = \vec S \cdot \left(
\vec h^{(0)} \tau_0 + \sum_{i = x,y,z} \vec h^{(i)} \tau_i
\right).
\eden
Here $\vec S=(s_x,s_y,s_z)/2$ is the spin operator, 
$\tau_0$ is the unit operator in valley space,
$s_i$ ($\tau_i$) denotes the Pauli matrices acting in spin (valley) space,
and the quantities $\vec h^{(0,x,y,z)}$ are different linear combinations
of the individual nuclear spin operators (see below). 
Equation \eqref{eq:hfsimple} should be contrasted with the widely used
$2\times 2$ hyperfine Hamiltonian $H_{\rm hf,GaAs}= \vec S\cdot \vec h$,
which describes the effect of the nuclear spin on a twofold degenerate level
in a GaAs QD, and incorporates only 
a single Overhauser field $\vec h$.
We estimate that the order of magnitude of the 
valley-conserving ($\sim \tau_0$) and valley-mixing parts of $H_{\rm hf}$ are the same.

\begin{figure}
\includegraphics[scale=0.5]{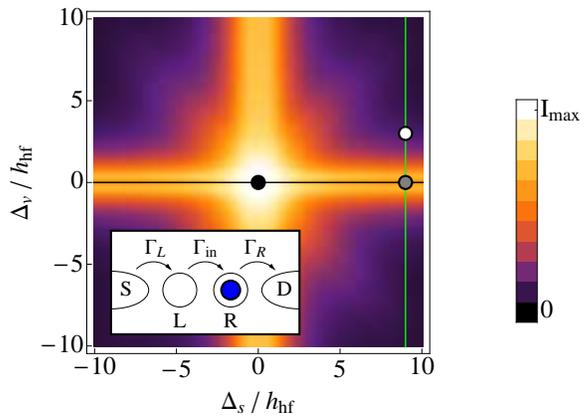}
\caption{\label{fig:density-plot}
(Color online)
Averaged leakage current $\langle I \rangle$
due to hyperfine interaction in a graphene DQD,
as a function of spin and valley splittings.
$I_{\rm max}\approx 0.34 e \Gamma_{\rm in}$.
Inset: transport through a DQD. Dot $R$ is always occupied by at least one electron.
}
\end{figure}

As a physical consequence of the valley coupling due to the 
HFI, we predict that the response of the
Pauli blockade leakage current through a carbon-based DQD to an
applied external magnetic field is remarkably different from the
case of GaAs DQDs.
In the Pauli or spin blockade regime \cite{Hanson-rmp}, transport from the source to the drain
through a serially coupled DQD occurs via the $(0,1)\rightarrow(1,1)\rightarrow(0,2)\rightarrow(0,1)$ 
cycle,
$(n_L,n_R)$ denoting the charge state with $n_L$ ($n_R$) electrons in the left (right) QD [see Fig. \ref{fig:density-plot} (inset)].
In a GaAs DQD, 
blocking of the current occurs 
%transport is blocked 
when there is at least one two-electron energy eigenstate in 
the (1,1) charge configuration having a spin wave function,
which is symmetric under particle 
exchange (i.e., a triplet). 
Due to HFI, this condition is achieved only in the
presence of an external magnetic field, which splits two triplet states
apart from the singlet state and prevents hyperfine-induced mixing of those \cite{Koppens-spinblockade,Jouravlev-spinblockade}. 
In carbon-based QDs with fourfold level degeneracy, 
the current is blocked only if there is at least one energy eigenstate in the (1,1)
charge configuration with a \emph{combined spin-valley wave function} which is symmetric
under particle exchange (`supertriplet').
In order to distinguish this effect from the spin blockade in conventional DQDs,
we call it the \emph{spin-valley blockade}.
We show that in contrast to GaAs, in 
%carbon-based
carbon
DQDs a spin (Zeeman) splitting is insufficient
to maintain the blockade in the presence of HFI, which, however, 
can be recovered by simultaneously introducing spin and valley splittings.

%\emph{Valley mixing due to hyperfine interaction.}
We consider the lowest-lying orbital level of an
electrostatically defined QD in monolayer graphene with a 
finite gap \cite{Recher-grapheneqds}.
(Nevertheless, the concepts and the formalism we use are
readily generalizable to other types of graphene or
CNT QDs.)
In the absence of nuclear spins and external magnetic field,
this level is fourfold (spin and valley) degenerate.
In the presence of an external magnetic field, both the spin
and valley degeneracy can be split: an in-plane magnetic
field causes only a spin splitting $\Delta_s=g_e\mu_B B$ 
via the Zeeman effect, whereas an out-of-plane ($z$) component 
introduces a valley splitting $\Delta_v(B_z)$ as well \cite{Recher-grapheneqds}.
We consider the regime, where the energy difference between the lowest-lying 
and the second orbital levels is much larger than $\Delta_s$ and $\Delta_v$.
We describe the system using the tight-binding (TB) model.
The TB wave functions, corresponding to the four sublevels
of the lowest-lying orbital level and characterized by the
spin and valley quantum numbers 
$s\in(\uparrow,\downarrow) \equiv (+,-)$
and 
$v\in(K,K') \equiv (+,-)$, are
$(\psi_{sv})_{l \sigma} = \sqrt{\Omega_{\rm cell}} e^{iv\vec K\cdot \vec r_{l \sigma}}
\Psi_\sigma^{(v)}(\vec r_{l \sigma}) \chi_s$.
Here $\sigma\in\{A,B\}$ is the sublattice index, $l$ is the unit cell index,
 $\Omega_{\rm cell}$ is the unit cell area, $\vec r_{l \sigma}$
is the position of the carbon atom on sublattice $\sigma$ in the $l$th unit cell,
and $\chi_+=(1,0)$ and $\chi_-=(0,1)$ are the two possible spin states. 
The four smoothly varying functions $\Psi_\sigma^{(v)}$ can be obtained by
solving the Dirac-like envelope function equation \cite{Recher-grapheneqds,Beenakker2009}.
The functions $\Psi_\sigma^{(v)}$ and $\psi_{sv}$ are normalized:
%Both the envelope functions and the TB wave functions fulfill the 
%normalization condition: 
$\int d^2\vec r \left( |\Psi_A^{(v)}(\vec r)|^2+|\Psi_B^{(v)}(\vec r)|^2 \right)=1$ 
and
$\sum_{l\sigma}|(\psi_{sv})_{l\sigma}|^2=1$.
%The TB wave functions can be used to express the 
%real-space wave functions $\Phi_{sv}(\vec r)$ as linear combinations of 
%atomic $p_z$ orbitals $\phi(\vec r)$ located on the carbon atoms:
%$
%\Phi_{sv}(\vec r) = \sum_{l\sigma} (\psi_{sv})_{l\sigma} \phi(\vec r-\vec r_{l\sigma})
%$.

%To describe the effect of the nuclear spins on the four sublevels of the
%QD level we start from the tight binding description of the HFI. 
The nuclear spin of the carbon atom on site $l\sigma$
is denoted by $\vec I_{l\sigma}$, being zero if the atom
is a $^{12}$C and a spin-1/2 operator if the atom is $^{13}$C. 
In graphene and CNTs, the HFI is known
to be short-range: its major contribution to 
the TB Hamiltonian is the on-site matrix element on the 
site of the nuclear spin \cite{Fischer-cnt}.
Therefore the elements of the TB Hamiltonian matrix,
describing the HFI, are
$
(H_{\rm hf,tb})_{l\sigma,l'\sigma'} = 
\delta_{ll'}\delta_{\sigma \sigma'} \vec S A \vec I_{l\sigma}
$,
where $A={\rm diag}(A_x,A_y,A_z)$ is a diagonal matrix.

The effective Hamiltonian describing the influence of the HFI on the
four sublevels is constructed by expressing $H_{\rm hf,tb}$ in the 
subspace spanned by the four TB wave functions $\psi_{sv}$.
The resulting matrix can be expressed in terms
of the spin and valley operators as shown in Eq. \eqref{eq:hfsimple}
using
\bean
 \label{eq:hyperfinefields}
   \vec h^{(k)} = \Omega_{\rm cell} A \sum_{l\sigma} \vec I_{l\sigma} F^{(k)}_{l\sigma} \qquad (k=0,x,y,z),
\eean
where
$F^{(0)}_{l\sigma} = \sum_{v}f^{(v)}_{l\sigma}/2$, 
$F^{(z)}_{l\sigma} = \sum_{v}vf^{(v)}_{l\sigma}/2$, 
$F^{(x/y)}_{l\sigma} = {\rm Re/Im}\left(e^{-2i\vec K \cdot \vec r_{l\sigma}} g_{l\sigma}\right)$, with
$f^{(v)}_{l\sigma}=|\Psi_\sigma^{(v)}(\vec r_{l\sigma})|^2$
and
$g_{l\sigma}=\Psi_\sigma^{(+)*} (\vec r_{l\sigma}) \Psi_\sigma^{(-)} (\vec r_{l\sigma})$.
Note that in the presence of time reversal symmetry
$\vec h^{(z)} = 0$.
Under normal conditions, the nuclear spins are completely
randomized by thermal fluctuations, which implies that
the components of the hyperfine fields have zero mean
and their variances are
\bean
 \label{eq:hyperfinevariances}
   \langle \left( h^{(k)}_j \right)^2 \rangle =
    \Omega_{\rm cell}^2 A_j^2 \frac{\nu}{4} 
    \sum_{l\sigma} \left(F^{(k)}_{l\sigma}\right)^2 \quad (k=0,x,y,z).
\eean
We have used $\langle (I_{l\sigma})^2_j\rangle=\nu / 4$
and $\nu$ denotes the abundance of $^{13}$C atoms in the QD.
Using the slowly varying nature of the envelope functions,
we find 
$\sum_{l\sigma} \left(F^{(x,y)}_{l\sigma}\right)^2 = \frac 1 2 \sum_{l\sigma} |g_{l\sigma}|^2$.
In the $N\to \infty$ limit 
($N$ is the number of atoms in the QD)
the distributions of the
hyperfine fields $h_j^{(k)}$ converge to Gaussians.
Furthermore, $h_j^{(x)}$ and $h_j^{(y)}$
become independent from 
$h_j^{(0)}$, $h_j^{(z)}$ and each other.

To determine the variances of the
hyperfine field precisely, one needs to know the
envelope functions $\Psi_\sigma^{(v)}$.
However, a simple estimation can be given using
the assumption $|\Psi_\sigma^{(v)}|^2 \approx 1/\Omega_{\rm cell} N$.
This choice satisfies the normalization condition, and it results in 
$\langle \left(h^{(0)}_j\right)^2 \rangle = A_j^2 \nu/4N$,
$\langle \left(h_j^{(x/y)}\right)^2 \rangle = A_j^2 \nu/8N$
and
$\langle \left(h_j^{(z)}\right)^2 \rangle=0$.
We will use these values in the following calculations.

The Hamiltonian $H_{\rm hf}$ in Eq. \eqref{eq:hfsimple}
shows that the HFI in carbon-based QDs
results both in spin and valley mixing,
and our estimates of the hyperfine field variances
suggest that the valley-conserving and valley-mixing
contributions in $H_{\rm hf}$ have the same order of magnitude. 
%Furthermore, according to our estimates of the hyperfine field variances,
%the characteristic energy scale of the valley mixing
%contributions $\vec h^{(x,y,z)}$ in $H_{\rm hf}$ is similar
%to that of the valley conserving contribution $\vec h^{(0)}$.
We emphasize that this result is due to the short-range nature of 
the HFI.
We anticipate a similar result in silicon QDs \cite{Shaji-siliconspinblockade,Culcer-silicondqds}, where
electrons also possess a valley degree of freedom.
%Our findings may have several important physical consequences, we highlight three of them here:
We highlight three possible physical consequences of the valley-mixing nature of $H_{\rm hf}$.
(i) 
HFI can be a source of valley relaxation and decoherence 
in `valleytronics' devices, such as valley filters, valley valves \cite{Rycerz-valleytronics}
or QD valley qubits \cite{Recher-graphenering}.
(ii)
Spin-orbit coupling in CNT QDs can split the fourfold
degenerate ground state dot level into two
 doublets 
($K \!\! \uparrow$, $K' \!\! \downarrow$ and $K' \!\! \uparrow$, $K \!\! \downarrow$)
at $B=0$ \cite{Kuemmeth-spinorbit-in-cnts}.
Spin-orbit coupling is stronger than the HFI; 
hence, a fully valley-conserving
$H_{\rm hf}$ would not be able to cause mixing within or between the
Kramers doublets. 
According to our result, $H_{\rm hf}$ 
is not valley-conserving and
causes mixing within the Kramers doublet.
This finding may facilitate the understanding 
of yet unexplained features of recent experiments
on $^{13}$C nanotubes \cite{Churchill-cntspinblockade,Churchill-13cntprl}.
(iii)
As we show below, the valley-mixing character of the HFI 
introduces remarkable features
in the behavior of the Pauli blockade leakage current through a 
carbon-based DQD.

%\emph{Pauli blockade.}
Our goal is to calculate the average leakage current through a
DQD in the Pauli blockade regime as a function of spin and valley splittings.
To this end, we have generalized the master equation formalism used in 
Ref. \onlinecite{Jouravlev-spinblockade}.
We focus on the parameter regime where the result Eq. (11) of 
Ref. \onlinecite{Jouravlev-spinblockade} is valid:
(i) The energy detuning $\Delta\equiv E(0,2)-E(1,1)$ 
between the (0,2) and (1,1) charge states
is much larger than the coherent tunneling strength between the two dots ($t$), 
the characteristic energy scale describing the HFI 
[$h_{\rm hf}=\sqrt{(\nu/4N) \sum_{j}A_j^2 }$], and 
the spin and valley splittings ($\Delta_s$, $\Delta_v$).
This condition suppresses the coherent tunneling between
the dots; therefore, the hybridization between $(1,1)$ and $(0,2)$
charge states is negligible.
(ii) The $(1,1)\to (0,2)$ transition is an
energetically downhill inelastic tunneling process,
characterized by the rate $\Gamma_{\rm in}$.
(iii) $t^2/|\Delta| \ll h_{\rm hf}$, which enables one to 
neglect the exchange splitting within (1,1) charge states.
(iv) $\Gamma_L \gg \Gamma_R \gg \Gamma_{\rm in}$, 
where the rate $\Gamma_L$ [$\Gamma_R$] describes the
$(0,1)\to(1,1)$ [$(0,2)\to(0,1)$] transition.
(v) $\Gamma_L < h_{\rm hf}$, which enables one to use a classical master
equation to describe the transport process.
Studying the Pauli blockade problem under the above specified conditions
is motivated by the fact that the theory for GaAs DQDs 
in this parameter regime has been successful in 
describing recent experimental results \cite{Koppens-spinblockade}.
We also adopt the constant interaction feature of the model used in 
Ref. \onlinecite{Jouravlev-spinblockade}.
Investigation of parameter regimes where the Coulomb interaction results in Wigner molecule formation 
\cite{Roy-coulombincnt,Wunsch-coulombincnt,Secchi-coulombincnt} 
is beyond the scope of this Rapid Communication.

\begin{figure}
\includegraphics[scale=0.95]{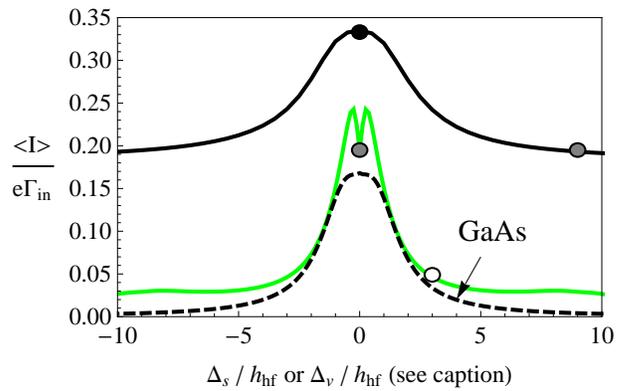}
\caption{\label{fig:jn-comparison}
(Color online)
Averaged leakage current $\langle I \rangle$ 
as a function of 
(upper solid line, black)
spin splitting $\Delta_s$ for zero valley splitting $\Delta_v = 0$,
(lower solid line, green)
valley splitting $\Delta_v$ for $\Delta_s = 9$,
(dashed)
$\Delta_s$ in a GaAs DQD
[dashed line based on Eq. (11) of Ref. \onlinecite{Jouravlev-spinblockade}].
}
\end{figure}

Our approach deviates from the one used in
Ref. \onlinecite{Jouravlev-spinblockade}
in the form of the single-particle Hamiltonians of the two QDs,
incorporating external magnetic field and HFI:
$H_{D} =  H_{D,{\rm magn}} + H_{D,{\rm hf}}$,
where 
$D\in\{L,R\}$,
$H_{D,{\rm magn}} = g_e\mu_B\vec S_D \cdot \vec B + \Delta_v(B_z) \tau_{z,D}/2$
and 
$H_{D,{\rm hf}} = 
\vec S_D \cdot \sum_{i = 0,x,y,z} \vec h_D^{(i)} \tau_{i,D}$.
According to our above analysis, we treat 
$\vec h_D^{(0,x,y,z)}$ as uncorrelated stationary random fields \cite{Jouravlev-spinblockade}
and use our previous estimates for their variances.
For our purposes, we can neglect the spin-anisotropy of the HFI
\cite{Yazyev-graphenehf,Fischer-cnt}, and use 
$A_j=A_{\rm iso}$ and therefore $h_{\rm hf} = A_{\rm iso}\sqrt{3\nu/4N}$.
Evaluating $h_{\rm hf}$ with the 
measured \cite{Churchill-cntspinblockade,Churchill-13cntprl} value 
$A_{\rm iso} = 100 \mu\rm{eV}$ and assuming $N=7.5\times10^4$ atoms in a QD 
we find $h_{\rm hf}\approx 30{\rm neV}$ ($300{\rm neV}$) for $^{13}$C abundance $\nu = 1\%$ ($100\%$);
using the theoretical estimate \cite{Yazyev-graphenehf,Fischer-cnt} 
$A_{\rm iso} = 1 \mu\rm{eV}$, 
we find $h_{\rm hf}\approx 0.3{\rm neV}$ ($3{\rm neV}$).
Under the above specified conditions, we set up a classical master equation for the
occupation probabilities of the (1,1) eigenstates of $H_L+H_R$.
By solving the master equation numerically, we calculate the stationary current $I$ for 
many realizations of the nuclear fields 
and average those to obtain $\langle I \rangle$.
In Fig. \ref{fig:density-plot}, we plot $\langle I \rangle$ as a function of spin and valley splittings
obtained by averaging for 200 random realizations 
of the hyperfine fields.
Two cuts along the horizontal (black) and vertical (green) lines of Fig. \ref{fig:density-plot} are shown
on Fig. \ref{fig:jn-comparison},
obtained by averaging over 3000 random realizations of
the nuclear fields. 
For comparison, in Fig. \ref{fig:jn-comparison} we also plotted the result corresponding to the case
of GaAs DQDs (dashed).

One of the characteristic features of the spin-valley blockade is
the cross-shaped pattern in the $\langle I (\Delta_s,\Delta_v) \rangle$
density plot in Fig. \ref{fig:density-plot}.
This pattern indicates that the current is strongly suppressed 
only if both the spin and valley splittings ($\Delta_s$ and $\Delta_v$) 
exceed the energy scale of the hyperfine coupling.
As mentioned earlier, a physical situation where $\Delta_s$ 
is finite but $\Delta_v=0$, is when 
an in-plane magnetic field is applied to the system.
This situation corresponds to the black line 
in Fig. \ref{fig:density-plot} and 
the solid black curve $\langle I (\Delta_s) \rangle$ in Fig. \ref{fig:jn-comparison}.
It is apparent from Fig. \ref{fig:jn-comparison} that
the current decreases with increasing $\Delta_s$;
but instead of approaching zero, it saturates to a finite value around $0.2 e \Gamma_{\rm in}$.
This is in stark contrast to the case of GaAs DQDs,
where $\Delta_s > h_{\rm hf}$
leads to a sharp decay of the current (dashed curve).

To present a qualitative interpretation of our results, 
we make use of the analogy between the spin and valley degrees of freedom.
We describe the states of the (1,1) charge configuration by using 
the simultaneous eigenbasis of the total spin operator $(\vec S_L+\vec S_R)^2$, 
the spin projection on the direction of the magnetic field 
$(\vec S_L+\vec S_R)\cdot \vec B/B$,
the total valley operator $(\vec \tau_L+\vec \tau_R)^2/4$,
and the $z$-component of the valley operator $(\tau_{z,L}+\tau_{z,R})/2$.
The corresponding quantum numbers are  
$s\in\{0,1\}$,  $m_s \in \{-s,\dots,s\}$, $v\in\{0,1\}$, and $m_v \in \{-v,\dots,
v\}$.
We denote these basis states with
$|s,m_s,v,m_v\rangle$.
These are eigenstates of the system Hamiltonian in 
the absence of HFI.
The combined spin-valley wave functions of the 
ten states fulfilling $s=v$ are supertriplets;
therefore, these states cannot be 
squeezed into a (0,2) charge configuration. 
In contrast, the spin-valley wave functions 
of the six states with $s\neq v$ are supersinglets;
hence, their transition to (0,2)
is allowed.

\begin{figure}[t]
\includegraphics[scale=0.5]{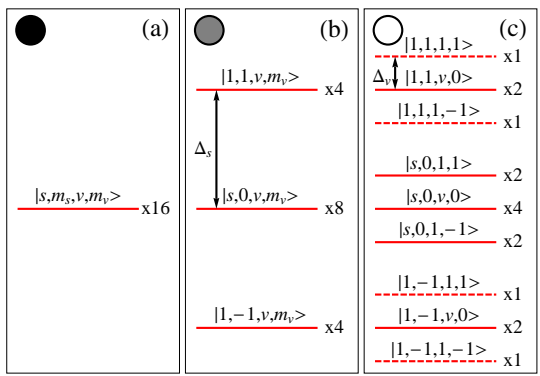}
\caption{\label{fig:energy-diagram}
(Color online)
Schematic energy diagrams of the (1,1) charge configuration
corresponding to the three highlighted points of Fig. \ref{fig:density-plot}.
The effect of hyperfine interaction is excluded.	
Dashed lines: transport-blocking supertriplet states.}
\end{figure}

The energy diagram of the 16 states of the
(1,1) charge configuration, 
corresponding to the three highlighted
points of Fig. \ref{fig:density-plot} are presented in Fig. \ref{fig:energy-diagram}.
Figure \ref{fig:energy-diagram}a shows the situation, where $\Delta_s=\Delta_v=0$.
In this case, there is a 16-fold degenerate level, 
and the HFI mixes supersinglet and supertriplet states effectively.
This results in a maximal current through the DQD.
The Zeeman effect splits the states with different $m_s$ quantum numbers (Fig. \ref{fig:energy-diagram}b)
and suppresses hyperfine-induced hybridization between them if $\Delta_s > h_{\rm hf}$,
which leads to a decrease in the leakage current.
However, the valley mixing contribution of the HFI
still induces strong mixing within the states with the same $m_s$.
This mixing prevents the appearance of ``pure'' supertriplet energy eigenstates
which would block the transport;
therefore, the current does not drop to zero.
As mentioned earlier, this behavior is in contrast to the case of 
GaAs DQDs.
Figure \ref{fig:energy-diagram}c
shows the energy diagram when both $\Delta_s$ and $\Delta_v$ are finite.
If those are larger than the HFI,
then the four supertriplet states
$|1,\pm 1,1,\pm 1\rangle$
become decoupled from supersinglets.
Thus, the system gets trapped whenever 
any of these four states is occupied during the transport process,
which results in a strong suppression of the current. 
Note that only two blocked states remain if $\Delta_s \approx \Delta_v$.
In that case, $|1,1,1,-1\rangle$ and $|1,-1,1,1\rangle$ become
degenerate with the fourfold degenerate $|s,0,v,0\rangle$
and mix with those due to HFI, slightly enhancing the current,
visible along the diagonal $|\Delta_s|=|\Delta_v|$ lines
in Fig. \ref{fig:density-plot}.

Another characteristic of the spin-valley blockade is the appearance of a dip 
in the green $\langle I (\Delta_v) \rangle $ curve in Fig. \ref{fig:jn-comparison} at $\Delta_v = 0$.
Similar dip structures have been predicted \cite{Qassemi-spinblockade,Danon-spinblockade}
and measured \cite{Koppens-spinblockade,Pfund-inasprl,Churchill-cntspinblockade} in
conventional semiconductors and they were attributed to various microscopic 
origins, including
cotunneling, spin-orbit interaction, and %the interplay of HFI and 
exchange coupling.
In our case, the dip has a different origin: 
it is due to the strong valley anisotropy of the HFI,
i.e., that in our above estimations $h_j^{(z)}$ vanishes and 
therefore $H_{\rm hf}$ does not include the $\tau_z$ operator.

%\emph{Conclusions.}
We have established the form of the Hamiltonian 
describing the effect of HFI on 
a fourfold degenerate energy level in a carbon-based QD.
We have found that the short range nature of the HFI
leads to a significant nuclear spin-electron valley coupling. 
We have calculated the effect of this interaction on the 
leakage current through a DQD in the Pauli blockade regime. 
Our findings may have profound consequences for both spin and valley
manipulation in carbon-based QDs. 

We thank K. Flensberg for a useful discussion,
and DFG for financial support within
SFB 767, SPP 1285, and FOR 912.

%===

\bibliography{spinblockade-letter}

\end{document}